\documentclass[12pt]{article}
\usepackage{axodraw,bbold}

\catcode`@=12
\begin{document}
\thispagestyle{empty}
\begin{flushright} 
UCRHEP-T412\\ 
June 2006\
\end{flushright}
\vspace{0.5in}
\begin{center}
{\LARGE	\bf Lepton Family Symmetry and\\ the Neutrino Mixing Matrix\\}
\vspace{1.5in}
{\bf Ernest Ma\\}
\vspace{0.2in}
{\sl Physics Department, University of California, Riverside, 
California 92521\\}
\vspace{1.5in}
\end{center}
\begin{abstract}\
I review some of the recent progress (up to September 2005) in applying 
non-Abelian discrete symmetries to the family structure of leptons, 
with particular emphasis on the tribimaximal mixing {\it ansatz} of Harrison, 
Perkins, and Scott.
\end{abstract}
\vspace{1.0in}
\centerline{---~~{\it Talk at Corfu2005}~~---}
\newpage
\baselineskip 24pt

\section{Introduction}

Using present data from neutrino oscillations, the $3 \times 3$ neutrino 
mixing matrix is largely determined, together with the two mass-squared 
differences \cite{data}.  In the Standard Model of particle interactions, 
there are 3 lepton families.  The charged-lepton mass matrix linking 
left-handed $(e, \mu, \tau)$ to their right-handed counterparts is in 
general arbitrary, but may always be diagonalized by 2 unitary 
transformations:
\begin{equation}
{\cal M}_l = U^l_L \pmatrix{m_e & 0 & 0 \cr 0 & m_\mu & 0 \cr 0 & 0 & m_\tau} 
(U^l_R)^\dagger.
\end{equation}
Similarly, the neutrino mass matrix may also be diagonalized by 2 unitary 
transformations if it is Dirac:
\begin{equation}
{\cal M}^D_\nu = U^\nu_L \pmatrix{m_1 & 0 & 0 \cr 0 & m_2 & 0 \cr 0 & 0 & 
m_3} (U^\nu_R)^\dagger,
\end{equation}
or by just 1 unitary transformation if it is Majorana:
\begin{equation}
{\cal M}^M_\nu = U^\nu_L \pmatrix{m_1 & 0 & 0 \cr 0 & m_2 & 0 \cr 0 & 0 & 
m_3} (U^\nu_L)^T.
\end{equation}
Notice that whereas the charged leptons have individual names, the 
neutrinos are only labeled as $1,2,3$, waiting to be named.
The observed neutrino mixing matrix is the mismatch between 
$U^l_L$ and $U^\nu_L$, i.e.
\begin{eqnarray}
U_{l\nu} = (U^l_L)^\dagger U^\nu_L \simeq \pmatrix{0.85 & 0.52 & 0.053 
\cr -0.33 & 0.62 & -0.72 \cr -0.40 & 0.59 & 0.70} \simeq \pmatrix{\sqrt{2/3} 
& 1/\sqrt{3} & 0 \cr -1/\sqrt{6} & 1/\sqrt{3} & -1/\sqrt{2} \cr -1/\sqrt{6} 
& 1/\sqrt{3} & 1/\sqrt{2}}. 
\end{eqnarray} 
This approximate pattern has been dubbed tribimaximal by Harrison, Perkins, 
and Scott \cite{hps}.  Notice that the 3 vertical columns are evocative  
of the mesons $(\eta_8,\eta_1,\pi^0)$ in their $SU(3)$ decompositions.\\

\noindent How can the HPS form of $U_{l\nu}$ be derived from a symmetry? The 
difficulty comes from the fact that any symmetry defined in the basis 
$(\nu_e,\nu_\mu,\nu_\tau)$ is automatically applicable to $(e,\mu,\tau)$ 
in the complete Lagrangian.  To do so, usually one assumes the canonical 
seesaw mechanism and studies the Majorana neutrino mass matrix
\begin{equation}
{\cal M}_\nu = -{\cal M}^D_\nu {\cal M}_N^{-1} ({\cal M}^D_\nu)^T
\end{equation}
in the basis where ${\cal M}_l$ is diagonal; but the symmetry apparent 
in ${\cal M}_\nu$ is often incompatible with a diagonal ${\cal M}_l$ with 
3 very different eigenvalues.\\

\noindent In this talk, I will discuss first the pitfall of $\mu 
\leftrightarrow \tau$ symmetry based on maximal $\nu_\mu-\nu_\tau$ mixing.  
I will show how it can be done properly with the permutation symmetry $S_3$.  
I will then spend most of the rest of my time on the tetrahedral symmetry 
$A_4$ and a little on the permutation symmetry $S_4$.  These are examples of 
how exact and approximate tribimaximal mixing may be obtained.

\section{Maximal $\nu_\mu-\nu_\tau$ Mixing}

Consider just 2 families. Suppose we want maximal $\nu_\mu-\nu_\tau$ 
mixing, then we should have
\begin{equation}
{\cal M}_\nu = \pmatrix{a & b \cr b & a} = {1 \over \sqrt2} \pmatrix{1 & -1 
\cr 1 & 1} \pmatrix{a+b & 0 \cr 0 & a-b} {1 \over \sqrt2} \pmatrix{1 & 1 \cr 
-1 & 1}.
\end{equation}
This seems to require the exchange symmetry $\nu_\mu \leftrightarrow 
\nu_\tau$, but since $(\nu_\mu,\mu)$ and $(\nu_\tau,\tau)$ are $SU(2)_L$ 
doublets, we must also have $\mu \leftrightarrow \tau$ exchange.  
Nevertheless, we still have the option of assigning $\mu^c$ and $\tau^c$.  
If $\mu^c \leftrightarrow \tau^c$ exchange is also assumed, then
\begin{equation}
{\cal M}_l = \pmatrix{A & B \cr B & A} = {1 \over \sqrt2} \pmatrix{1 & -1 
\cr 1 & 1} \pmatrix{A+B & 0 \cr 0 & A-B} {1 \over \sqrt2} \pmatrix{1 & 1 \cr 
-1 & 1}.
\end{equation}
Hence $U_{l\nu} = (U^l_L)^\dagger U^\nu_L = 1$ and there is no mixing.
If $\mu^c$ and $\tau^c$ do not transform under this exchange, then
\begin{equation}
{\cal M}_l = \pmatrix{A & B \cr A & B} = {1 \over \sqrt2} \pmatrix{1 & -1 
\cr 1 & 1} \pmatrix{\sqrt{2(A^2+B^2)} & 0 \cr 0 & 0} \pmatrix{c & s \cr 
-s & c},
\end{equation}
where $c=A/\sqrt{A^2+B^2}$, $s=B/\sqrt{A^2+B^2}$.  Again $U_{l\nu} = 
(U^l_L)^\dagger U^\nu_L = 1$.

\section{Permutation Symmetry $S_3$}

To overcome the difficulty of obtaining maximal $\nu_\mu-\nu_\tau$ mixing, 
consider the non-Abelian discrete symmetry $S_3$, i.e. the 
permutation group of 3 objects, which is also the symmetry group of the 
equilateral triangle.  It has 6 elements divided into 3 equivalence 
classes, with the irreducible representations \underline{1}, 
\underline{1}$'$, and \underline{2}. Let
\begin{equation}
\omega = \exp \left( {2\pi i \over 3} \right) = -{1 \over 2} + i 
{\sqrt 3 \over 2},
\end{equation}
then the 6 matrices of the \underline{2} representation may be chosen as 
follows.
\begin{eqnarray}
C_1: \pmatrix{1 & 0 \cr 0 & 1}, ~~~ C_2: \pmatrix{\omega & 0 \cr 0 & 
\omega^2}, \pmatrix{\omega^2 & 0 \cr 0 & \omega}, ~~~ 
C_3: \pmatrix{0 & 1 \cr 1 & 0}, \pmatrix{0 & \omega \cr \omega^2 & 0}, 
\pmatrix{0 & \omega^2 \cr \omega & 0},
\end{eqnarray}
where $C_i$ refer to the 3 equivalence classes in the character table shown.
\begin{table}[htb]
\centering
\caption{Character table of $S_3$.} 
\begin{tabular}{cccccc}
\hline
class&$n$&$h$&$\chi_1$&$\chi_{1'}$&$\chi_2$\\
\hline
$C_1$&1&1&1&1&2\\
$C_2$&2&3&1&1&--1\\
$C_3$&3&2&1&--1&0\\
\hline
\end{tabular}
\end{table}
The fundamental multiplication rule is then
\begin{equation}
\underline{2} \times \underline{2} = \underline{1}(12+21) + 
\underline{1}'(12-21) + \underline{2}(22,11). 
\end{equation}
Let $(\nu_i,l_i) \sim \underline{2}$, $l^c_i \sim \underline{2}$, 
$(\phi^0_1,\phi^-_1) \sim \underline{1}$, $(\phi^0_2,\phi^-_2) \sim 
\underline{1}'$, then
\begin{equation}
{\cal M}_l = \pmatrix{0 & fv_1+f'v_2 \cr fv_1-f'v_2 & 0} = \pmatrix{m_\mu & 0 
\cr 0 & m_\tau} \pmatrix{0 & 1 \cr 1 & 0}.
\end{equation}
Let $\xi_i = (\xi_i^{++},\xi_i^+,\xi_i^0) \sim \underline{2}$ and 
$\xi_0 \sim \underline{1}$, then
\begin{equation}
{\cal M}_\nu = \pmatrix{hu_1 & h_0u_0 \cr h_0u_0 & hu_2} = \pmatrix
{a & b \cr b & a}
\end{equation}
for $u_1=u_2$.  Thus
\begin{equation}
U_{l\nu} = (U^l_L)^\dagger U^\nu_L = {1 \over \sqrt 2} \pmatrix{1 & -1 \cr 
1 & 1},
\end{equation}
i.e. maximal $\nu_\mu-\nu_\tau$ mixing may be achieved, despite having 
a diagonal ${\cal M}_l$ with $m_\mu \neq m_\tau$.

\section{Tetrahedral Symmetry $A_4$}

For 3 families, we should look for a group with a \underline{3} 
representation, the simplest of which is $A_4$, the group of the even 
permutation of 4 objects, which is also the symmetry group of the 
tetrahedron.
\begin{table}[htb]
\centering
\caption{Character table of $A_4$.} 
\begin{tabular}{ccccccc}
\hline
class&$n$&$h$&$\chi_1$&$\chi_{1'}$&$\chi_{1''}$&$\chi_3$\\
\hline
$C_1$&1&1&1&1&1&3\\
$C_2$&4&3&1&$\omega$&$\omega^2$&0\\
$C_3$&4&3&1&$\omega^2$&$\omega$&0\\
$C_4$&3&2&1&1&1&--1\\
\hline
\end{tabular}
\end{table}

\noindent The fundamental multiplication rule is
\begin{eqnarray}
\underline{3} \times \underline{3} &=& \underline{1}(11+22+33) + 
\underline{1}'(11+\omega^222+\omega33) + \underline{1}''
(11+\omega22+\omega^233) \nonumber \\ &+& \underline{3}(23,31,12) + 
\underline{3}(32,13,21).
\end{eqnarray} 
Note that $\underline{3} \times \underline{3} \times \underline{3} = 
\underline{1}$ is possible in $A_4$, i.e. $a_1 b_2 c_3 +$ permutations, 
and $\underline{2} \times \underline{2} \times \underline{2} = \underline{1}$ 
is possible in $S_3$, i.e. $a_1 b_1 c_1 + a_2 b_2 c_2$.

\begin{table}[htb]
\centering
\caption{Perfect geometric solids.} 
\begin{tabular}{ccccc}
\hline
solid&faces&vertices&Plato&group\\
\hline
tetrahedron&4&4&fire&$A_4$\\
octahedron&8&6&air&$S_4$\\
icosahedron&20&12&water&$A_5$\\
hexahedron&6&8&earth&$S_4$\\
dodecahedron&12&20&quintessence&$A_5$\\
\hline
\end{tabular}
\end{table}

\noindent The tetrahedron is one of five perfect geometric solids known to the 
ancient Greeks.  In order to match them to the 4 elements (fire, air, 
water, and earth) already known, Plato invented a fifth (quintessence) 
as that which pervades the cosmos and presumably holds it together.
Since a cube (hexahedron) may be embedded inside an octahedron and vice versa, 
the two must have the same group structure and are thus dual to each other. 
The same holds for the icosahedron and dodecahedron.  The tetrahedron is 
self-dual. Compare this first theory of everything to today's contender, 
i.e. string theory.  (A) There are 5 consistent string theories in 10 
dimensions. (B) Type I is dual to Heterotic $SO(32)$, Type IIA 
is dual to Heterotic $E_8 \times E_8$, and Type IIB is self-dual.

\subsection{Exact HPS}

Following the original papers \cite{mr01,bmv03} on $A_4$, let $(\nu_i,l_i) 
\sim \underline{3}$, but $l^c_i \sim \underline{1}, \underline{1}', 
\underline{1}''$, then with $(\phi_i^0,\phi_i^-) \sim \underline{3}$,
\begin{equation}
{\cal M}_l = \pmatrix{h_1v_1 & h_2v_1 & h_3v_1 \cr h_1 v_2 & h_2 v_2 \omega 
& h_3 v_2 \omega^2 \cr h_1 v_3 & h_2 v_3 \omega^2 & h_3 v_3 \omega} = 
{1 \over \sqrt 3} \pmatrix{1 & 1 & 1 \cr 1 & \omega & \omega^2 
\cr 1 & \omega^2 & \omega} \pmatrix{h_1 & 0 & 0 \cr 0 & h_2 & 0 \cr 0 & 0 
& h_3} \sqrt{3} v,
\end{equation}
for $v_1=v_2=v_3=v$.
Let $\xi_1 \sim \underline{1}$, $\xi_2 \sim \underline{1}'$, $\xi_3 \sim 
\underline{1}''$, $\xi_{4,5,6} \sim \underline{3}$, with $\langle \xi_5 
\rangle = \langle \xi_6 \rangle = 0$, then \cite{m04}
\begin{equation}
{\cal M}_\nu = \pmatrix{a+b+c & 0 & 0 \cr 0 & a + b\omega + c\omega^2 & d \cr 
0 & d & a+b\omega^2+c\omega}.
\end{equation}
In the $(\nu_e,\nu_\mu,\nu_\tau)$ basis, it becomes
\begin{equation}
{\cal M}^{(e,\mu,\tau)}_\nu = \pmatrix{a+2d/3 & b-d/3 & c-d/3 \cr b-d/3 & 
c+2d/3 & a-d/3 \cr c-d/3 & a-d/3 & b+2d/3}.
\end{equation}
If $b=c$, then the eigenvalues of this matrix are simply
\begin{equation}
m_1=a-b+d, ~~~ m_2=a+2b, ~~~ m_3=-a+b+d,
\end{equation}
and
\begin{equation}
U_{l\nu} = \pmatrix{\sqrt{2/3} 
& 1/\sqrt{3} & 0 \cr -1/\sqrt{6} & 1/\sqrt{3} & -1/\sqrt{2} \cr -1/\sqrt{6} 
& 1/\sqrt{3} & 1/\sqrt{2}},
\end{equation}
i.e. tribimaximal mixing is obtained as desired.  If $b \neq c$, then 
$U_{e3} \neq 0$, and $|U_{e3}| < 0.16$ implies $0.5 < \tan^2 \theta_{12} 
< 0.52$, whereas experimentally, $\tan^2 \theta_{12} = 0.45 \pm 0.05$.\\

\noindent The above pattern involves 4 parameters $(a,b,c,d)$.  If a model 
can be constructed for which $b=c$ naturally, then the HPS {\it ansatz} of 
tribimaximal mixing will be realized.  Of course, the three masses are 
not predicted, as shown in Eq.~(19).  If $b \neq 0$ and $c \neq 0$, it is 
difficult, if not impossible, to find an auxiliary symmetry which will 
enforce their equality.  On the other hand, they can both be zero, and thus 
equal to each other, if $\xi_2$ and $\xi_3$ are absent in the above. This 
is the essence of how the problem is first solved by Altarelli and 
Feruglio \cite{af05}.  In that case,
\begin{equation}
m_1=a+d, ~~~ m_2=a, ~~~ m_3=-a+d.
\end{equation}
The requirement $\Delta m^2_{12} \simeq \Delta m^2_{sol} << \Delta m^2_{atm} 
\simeq \Delta m^2_{23}$ implies
\begin{equation}
|d| \simeq -2|a|\cos \phi, ~~~ |m_{1,2}|^2 \simeq |a|^2, ~~~ |m_3|^2 \simeq 
|a|^2 (1+8\cos^2 \phi),
\end{equation}
i.e. normal ordering of neutrino masses with the sum rule \cite{m05}
\begin{equation}
|m_{\nu_e}|^2 \simeq |m_{ee}|^2 + \Delta m^2_{atm}/9,
\end{equation}
where $|m_{\nu_e}|$ is the kinematic $\nu_e$ mass measured in beta decay 
and $|m_{ee}|$ is the effective Majorana neutrino mass measured in 
neutrinoless double beta decay.\\

\noindent Another 2-parameter tribimaximal scenario \cite{m05} is to choose 
$a=0$, $b=c$.  In that case,
\begin{equation}
m_1=-b+d, ~~~ m_2=2b, ~~~ m_3=b+d.
\end{equation}
Here both normal and inverted ordering of neutrino 
masses are possible with the sum rule
\begin{equation}
|m_{\nu_e}|^2 \simeq 3|m_{ee}|^2 - (2/3) \Delta m^2_{atm},
\end{equation}
More recently, exact HPS was obtained by Babu and He \cite{bh05} with $A_4$, 
using the canonical seesaw mechanism.  Their solution may be considered 
the ``inverse'' of Ref.~\cite{af05}.  Another example of exact HPS was 
obtained by Grimus and Lavoura \cite{gl05} with $S_3$ plus 1 commuting 
and 6 noncommuting $Z_2$ symmetries.

\subsection{Approximate HPS}

An alternative $A_4$ assignment \cite{hmvv05} is to let $(\nu_i,l_i), l^c_i 
\sim \underline{3}$ with $(\phi^0_i,\phi^-_i) \sim \underline{1}, 
\underline{1}', \underline{1}''$, then ${\cal M}_l$ is diagonal with
\begin{equation}
\pmatrix{m_e \cr m_\mu \cr m_\tau} = \pmatrix{1 & 1 & 1 \cr 1 & \omega 
& \omega^2 \cr 1 & \omega^2 & \omega} \pmatrix{h_1v_1 \cr h_2v_2 \cr h_3v_3}.
\end{equation}
For the neutrino mass matrix, let $\xi_1 \sim \underline{1}$, $\xi_2 \sim 
\underline{1}'$, $\xi_3 \sim \underline{1}''$, $\xi_{4,5,6} \sim 
\underline{3}$, with $\langle \xi_4 \rangle = \langle \xi_5 \rangle = 
\langle \xi_6 \rangle$, then
\begin{equation}
{\cal M}_\nu = \pmatrix{a+b+c & d & d \cr d & a+b\omega+c\omega^2 & d \cr 
d & d & a+b\omega^2+c\omega}.
\end{equation}
Let $b=c$ and rotate to the basis $[\nu_e,(\nu_\mu+\nu_\tau)/\sqrt 2, 
(-\nu_\mu+\nu_\tau)/\sqrt 2]$, then
\begin{equation}
{\cal M}_\nu = \pmatrix{a+2b & \sqrt 2 d & 0 \cr \sqrt 2 d & a-b+d & 0 \cr 
0 & 0 & a-b-d},
\end{equation}
i.e. maximal $\nu_\mu - \nu_\tau$ mixing and $U_{e3}=0$.  The solar mixing 
angle is now given by $\tan 2 \theta_{12} = 2\sqrt 2 d/(d-3b)$.  For 
$b << d$, $\tan 2 \theta_{12} \to 2\sqrt 2$, i.e. $\tan^2 \theta_{12} \to 
1/2$, but $\Delta m^2_{sol} << \Delta m^2_{atm}$ implies $2a+b+d \to 0$, so 
that $\Delta m^2_{atm} \to 6bd \to 0$ as well.  Therefore, $b \neq 0$ is 
required, and $\tan^2 \theta_{12} \neq 1/2$, but should be close to it, 
because $b=0$ enhances the symmetry of ${\cal M}_\nu$ from $Z_2$ to $S_3$. 
Here $\tan^2 \theta_{12} < 1/2$ implies inverted ordering and 
$\tan^2 \theta_{12} > 1/2$ implies normal ordering.

\section{Permutation Symmetry $S_4$}

In the above application of $A_4$, approximate tribimaximal mixing involves 
the {\it ad hoc} assumption $b=c$.  This problem is overcome by using $S_4$ in 
a supersymmetric seesaw model proposed recently \cite{s4}, yielding the 
result
\begin{equation}
{\cal M}_\nu = \pmatrix{a+2b & c & c \cr c & a-b & d \cr 
c & d & a-b}.
\end{equation}
Here $b=0$ and $c=d$ are related limits.  The permutation group of 4 
objects is $S_4$.  It contains both $S_3$ and $A_4$.  It is also the 
symmetry group of the hexahedron (cube) and the octahedron.
\begin{table}[htb]
\centering
\caption{Character table of $S_4$.} 
\begin{tabular}{cccccccc}
\hline
class&$n$&$h$&$\chi_1$&$\chi_{1'}$&$\chi_2$&$\chi_3$&$\chi_{3'}$\\
\hline
$C_1$&1&1&1&1&2&3&3\\
$C_2$&3&2&1&1&2&--1&--1\\
$C_3$&8&3&1&1&--1&0&0\\
$C_4$&6&4&1&--1&0&--1&1\\
$C_5$&6&2&1&--1&0&1&--1\\
\hline
\end{tabular}
\end{table}

\noindent The fundamental multiplication rules are
\begin{eqnarray}
\underline{3} \times \underline{3} &=& \underline{1}(11+22+33) + 
\underline{2}(11+\omega^222+\omega33,11+\omega22+\omega^233) \nonumber \\ 
&+& \underline{3}(23+32,31+13,12+21) + \underline{3}'(23-32,31-13,12-21),\\
\underline{3}' \times \underline{3}' &=& \underline{1} + 
\underline{2} + \underline{3}_S + \underline{3}'_A,\\
\underline{3} \times \underline{3}' &=& \underline{1}' + 
\underline{2} + \underline{3}'_S + \underline{3}_A.
\end{eqnarray} 
Note that both $\underline{3} \times \underline{3} \times \underline{3} = 
\underline{1}$ and $\underline{2} \times \underline{2} \times \underline{2} 
= \underline{1}$ are possible in $S_4$.
Let $(\nu_i,l_i),l^c_i,N_i \sim \underline{3}$ under $S_4$.  Assume singlet 
superfields $\sigma_{1,2,3} \sim \underline{3}$ and $\zeta_{1,2} \sim 
\underline{2}$, then
\begin{equation}
{\cal M}_N = \pmatrix{A+f(\langle \zeta_2 \rangle + \langle \zeta_1 \rangle) 
& h \langle \sigma_3 \rangle & h \langle \sigma_2 \rangle \cr 
h \langle \sigma_3 \rangle & A + f(\langle \zeta_2 \rangle \omega + 
\langle \zeta_1 \rangle \omega^2) & h \langle \sigma_1 \rangle \cr 
h \langle \sigma_2 \rangle & h \langle \sigma_1 \rangle & A + 
f(\langle \zeta_2 \rangle \omega^2 + \langle \zeta_1 \rangle \omega)}.
\end{equation}
The most general $S_4$-invariant superpotential of $\sigma$ and $\zeta$ is 
given by
\begin{eqnarray}
W &=& M(\sigma_1 \sigma_1 + \sigma_2 \sigma_2 + \sigma_3 \sigma_3) + 
\lambda \sigma_1 \sigma_2 \sigma_3 + m \zeta_1 \zeta_2 + \rho(\zeta_1 
\zeta_1 \zeta_1 + \zeta_2 \zeta_2 \zeta_2) \nonumber \\ 
&+& \kappa[(\sigma_1 \sigma_1 + \sigma_2 \sigma_2 \omega + \sigma_3 \sigma_3 
\omega^2) \zeta_2 + (\sigma_1 \sigma_1 + \sigma_2 \sigma_2 \omega^2 + 
\sigma_3 \sigma_3 \omega) \zeta_1].
\end{eqnarray}
The resulting scalar potential has a minimum at $V=0$ (thus preserving 
supersymmetry) only if $\langle \zeta_1 \rangle = \langle \zeta_2 \rangle$ 
and $\langle \sigma_2 \rangle = \langle \sigma_3 \rangle$, so that 
\begin{equation}
{\cal M}_N = \pmatrix{A+2B & C & C \cr C & A-B & D \cr 
C & D & A-B}.
\end{equation}
To obtain a diagonal ${\cal M}_l$, choose $\phi^l_{1,2,3} \sim 
\underline{1} + \underline{2}$.  To obtain a Dirac ${\cal M}_{\nu N}$ 
proportional to the identity, choose $\phi^N_{1,2,3} \sim \underline{1} 
+ \underline{2}$ as well, but with zero vacuum expectation value for 
$\phi^N_{2,3}$.  This allows ${\cal M}_\nu$ to have the form of Eq.~(29), 
and thus approximate tribimaximal mixing.

\section{Conclusion and Outlook}

Since my talk on finite groups in Dubrovnik exactly two years ago (which was 
itself exactly two years after my talk at the Gran Sasso Laboratory on that 
fateful day), much progress has been made.\\
 
\noindent With the application of the non-Abelian discrete symmetry $A_4$, 
a plausible theoretical understanding of the HPS form of the neutrino mixing 
matrix has been achieved, i.e. $\tan^2 \theta_{23} = 1$, $\tan^2 \theta_{12} 
= 1/2$, $\tan^2 \theta_{13} = 0$.\\

\noindent Another possibility is that $\tan^2 \theta_{12}$ is not 1/2, but 
close to it. This has theoretical support in an alternative version of $A_4$, 
but is much more natural in $S_4$.\\

\noindent In the future, this approach to lepton family symmetry should be 
extended to include quarks, perhaps together in a consistent overall theory, 
such as $SU(3)^3$ finite unification \cite{mmz04}.

\section*{Acknowledgement}
I thank George Zoupanos and the other organizers of Corfu2005 for 
their great hospitality and a stimulating conference.  This work is supported 
by the EPEAEK programme ``Pythagoras II'' and co-funded by the European 
Union (75\%) and the Hellenic state (25\%).


\begin{thebibliography}{99}
\bibitem{data} See for example S. Goswami, in 2005 Lepton Photon Symposium, 
Uppsala, Sweden (to be published).
\bibitem{hps} P. F. Harrison, D. H. Perkins, and W. G. Scott, Phys. Lett. 
{\bf B530}, 167 (2002).  See also X.-G. He and A. Zee, Phys. Lett. {\bf B560}, 
87 (2003).
\bibitem{mr01} E. Ma and G. Rajasekaran, Phys. Rev. {\bf D64}, 113012 (2001).
\bibitem{bmv03} K. S. Babu, E. Ma, and J. W. F. Valle, Phys. Lett. 
{\bf B552}, 207 (2003).
\bibitem{m04} E. Ma, Phys. Rev. {\bf D70}, 031901(R) (2004).
\bibitem{af05} G. Altarelli and F. Feruglio, Nucl. Phys. {\bf B720}, 64 
(2005).
\bibitem{m05} E. Ma, Phys. Rev. {\bf D72}, 037301 (2005).
\bibitem{bh05} K. S. Babu and X.-G. He, hep-ph/0507217.
\bibitem{gl05} W. Grimus and L. Lavoura, JHEP {\bf 0601}, 018 (2006).
\bibitem{hmvv05} M. Hirsch, E. Ma, A. Villanova del Moral, and J. W. F. 
Valle, Phys. Rev. {\bf D72}, 091301(R) (2005); Erratum-ibid. {\bf D72}, 
119904 (2005).
\bibitem{s4} E. Ma, Phys. Lett. {\bf B632}, 352 (2006).
\bibitem{mmz04} E. Ma, M. Mondragon, and G. Zoupanos, JHEP {\bf 0412}, 
026 (2004).
\end{thebibliography}
\end{document}